# From Nobel Prize to Project Management: Getting Risks Right

by

Bent Flyvbjerg, Ph.D. & Dr. Techn.
Aalborg University, Denmark

**Abstract**

A major source of risk in project management is inaccurate forecasts of project costs, demand, and other impacts. The paper presents a promising new approach to mitigating such risk, based on theories of decision making under uncertainty which won the 2002 Nobel prize in economics. First, the paper documents inaccuracy and risk in project management. Second, it explains inaccuracy in terms of optimism bias and strategic misrepresentation. Third, the theoretical basis is presented for a promising new method called "reference class forecasting," which achieves accuracy by basing forecasts on actual performance in a reference class of comparable projects and thereby bypassing both optimism bias and strategic misrepresentation. Fourth, the paper presents the first instance of practical reference class forecasting, which concerns cost forecasts for large transportation infrastructure projects. Finally, potentials for and barriers to reference class forecasting are assessed.

**The American Planning Association Endorses Reference Class Forecasting**

In April 2005, based on a study of inaccuracy in demand forecasts for public works projects by Flyvbjerg, Holm, and Buhl (2005), the American Planning Association (APA) officially endorsed a promising new forecasting method called "reference class forecasting" and made the strong recommendation that planners should never rely solely on conventional forecasting techniques when making forecasts:

> "APA encourages planners to use reference class forecasting in addition to traditional methods as a way to improve accuracy. The reference class forecasting method is beneficial for non-routine projects ... Planners should never rely solely on civil engineering technology as a way to generate project forecasts" (the American Planning Association 2005).

Reference class forecasting is based on theories of decision making under uncertainty that won Princeton psychologist Daniel Kahneman the Nobel prize in economics in 2002 (Kahneman and Tversky 1979a, b; Kahneman 1994). Reference class forecasting promises more accuracy in forecasts by taking a so-called "outside view" on prospects being forecasted, while conventional forecasting takes an inside view. The outside view on a given project is based on knowledge about actual performance in a reference class of comparable projects.

Where Flyvbjerg, Holm, and Buhl (2005) briefly outlined the idea of reference class forecasting, this paper presents the first instance of reference class forecasting in practical project management. The emphasis will be on transportation project management, because this is where the first instance of reference class forecasting occurred. It should be mentioned at the outset, however, that comparative research shows that the problems, causes, and cures identified for transportation apply to a wide range of other project types including concert halls, museums, sports arenas, exhibit and convention centers, urban renewal, power plants, dams, water projects, IT systems, oil and gas extraction projects, aerospace projects, new production plants, and the development of new products and new markets (Altshuler and Luberoff 2003; Flyvbjerg, Bruzelius, and Rothengatter 2003: 18-19; Flyvbjerg, Holm, and Buhl 2002: 286; Flyvbjerg 2005).

## Inaccuracy in Forecasts

Forecasts of cost, demand, and other impacts of planned projects have remained constantly and remarkably inaccurate for decades. No improvement in forecasting accuracy seems to have taken place, despite all claims of improved forecasting models, better data, etc. (Flyvbjerg, Bruzelius, and Rothengatter 2003; Flyvbjerg, Holm, and Buhl 2002, 2005). For transportation infrastructure projects inaccuracy in cost forecasts in constant prices is on average 44.7% for rail, 33.8% for bridges and tunnels, and 20.4% for roads (see Table 1).[i] For the 70-year period for which cost data are available, accuracy in cost forecasts has not improved. Average inaccuracy for rail passenger forecasts is –51.4%, with 84% of all rail projects being wrong by more than ±20%. For roads, average inaccuracy in traffic forecasts is 9.5%, with half of all road forecasts being wrong by more than ±20% (see Table 2). For the 30-year period for which demand data are available, accuracy in rail and road traffic forecasts has not improved.

When cost and demand forecasts are combined, for instance in the cost-benefit analyses that are typically used to justify large transportation infrastructure investments, the consequence is inaccuracy to the second degree. Benefit-cost ratios are often wrong, not only by a few percent but by several factors. This is especially the case for rail projects (Flyvbjerg, Bruzelius, and Rothengatter 2003, 37-41). As a consequence, estimates of viability are often misleading, as are socioeconomic and environmental appraisals, the accuracy of which are all heavily dependent on demand and cost forecasts. These results point to a significant problem in transportation project management: More often than not the information that managers use to decide whether to invest in new projects is highly inaccurate and biased making projects highly risky. Comparative studies show that transportation projects are no worse than other project types in this respect (Flyvbjerg, Bruzelius, and Rothengatter 2003).

**Explaining Inaccuracy**

Flyvbjerg, Holm, and Buhl (2002, 2004, 2005) and Flyvbjerg and Cowi (2004) tested technical, psychological, and political-economic explanations for inaccuracy in forecasting. Technical explanations are most common in the literature and they explain inaccuracy in terms of unreliable or outdated data and the use of inappropriate forecasting models (Vanston and Vanston 2004: 33). However, when such explanations are put to empirical test they do not account well for the available data. First, if technical explanations were valid one would expect the distribution of inaccuracies to be normal or near-normal with an average near zero. Actual distributions of inaccuracies are consistently and significantly non-normal with averages that are significantly different from zero. Thus the problem is bias and not inaccuracy as such. Second, if imperfect data and models were main explanations of inaccuracies, one would expect an improvement in accuracy over time, since in a professional setting errors and their sources would be recognized and addressed, for instance through referee processes with scholarly journals and similar critical expert reviews. Undoubtedly, substantial resources have been spent over several decades on improving data and forecasting models. Nevertheless, this has had no effect on the accuracy of forecasts as demonstrated above. This indicates that something other than poor data and models is at play in generating inaccurate forecasts. This finding has been corroborated by interviews with forecasters (Flyvbjerg and Cowi 2004; Flyvbjerg and Lovallo in progress; Wachs 1990).

Psychological and political explanations better account for inaccurate forecasts. Psychological explanations account for inaccuracy in terms of optimism bias, that is, a cognitive predisposition found with most people to judge future events in a more positive light than is warranted by actual experience. Political explanations, on the other hand, explain inaccuracy in terms of strategic misrepresentation. Here, when forecasting the outcomes of projects, forecasters and managers deliberately and strategically overestimate benefits and underestimate costs in order to increase the likelihood that it is their projects, and not the competition's, that gain approval and funding. Strategic misrepresentation can be traced to political and organizational pressures, for instance competition for scarce funds or jockeying for position. Optimism bias and strategic misrepresentation are both deception, but where the latter is intentional, i.e., lying, the first is not, optimism bias is self-deception. Although the two types of explanation are different, the result is the same: inaccurate forecasts and inflated benefit-cost ratios. However, the cures to optimism bias are different from the cures to strategic misrepresentation, as we will see below.

Explanations of inaccuracy in terms of optimism bias have been developed by Kahneman and Tversky (1979a) and Lovallo and Kahneman (2003). Explanations in terms of strategic misrepresentation have been set forth by Wachs (1989,1990) and Flyvbjerg, Holm, and Buhl (2002, 2005). As illustrated schematically in Figure 1, explanations in terms of optimism bias have their relative merit in situations where political and organizational pressures are absent or low, whereas such explanations hold less power in situations where political pressures are high. Conversely, explanations in terms of strategic misrepresentation have their relative merit where political and organizational pressures are high, while they become immaterial when such pressures are not present. Thus, rather than compete, the two types of explanation complement each other: one is strong where the other is weak, and both explanations are necessary to understand the phenomenon at hand--the pervasiveness of inaccuracy in forecasting--and how to curb it.

In what follows a forecasting method called "reference class forecasting" is presented, which bypasses human bias--including optimism bias and strategic misrepresentation--by cutting directly to outcomes. In experimental research carried out by Daniel Kahneman and others, this method has been demonstrated to be more accurate than conventional forecasting methods (Kahneman and Tversky 1979a, 1979 b; Kahneman 1994; Lovallo and Kahneman 2003). First the theoretical and methodological foundations for reference class forecasting are explained, then the first instance of reference class forecasting in project management is presented.

## The Planning Fallacy and Reference Class Forecasting

The theoretical and methodological foundations of reference class forecasting were first described by Kahneman and Tversky (1979b) and later by Lovallo and Kahneman (2003). Reference class forecasting was originally developed to compensate for the type of cognitive bias that Kahneman and Tversky found in their work on decision making under uncertainty, which won Kahneman the Nobel prize in economics 2002 (Kahneman 1994; Kahneman and Tversky 1979a). This work showed that errors of judgment are often systematic and predictable rather than random, manifesting bias rather than confusion, and that any corrective prescription should reflect this. They also found that many errors of judgment are shared by experts and laypeople alike. Finally they found that errors remain compelling even when one is fully aware of their nature. Thus awareness of a perceptual or cognitive illusion does not by itself produce a more accurate perception of reality, according to Kahneman and Tversky (1979b: 314). Awareness may, however, enable one to identify situations in which the normal faith in one's impressions must be suspended and in which judgment should be controlled by a more critical evaluation of the evidence. Reference class forecasting is a method for such critical evaluation. Human judgment, including forecasts, are biased. Reference class forecasting is a method for debiasing forecasts.

Kahneman and Tversky (1979a, b) found human judgment to be generally optimistic due to overconfidence and insufficient regard to distributional information. Thus people will underestimate the costs, completion times, and risks of planned actions, whereas they will overestimate the benefits of the same actions. Lovallo and Kahneman (2003: 58) call such common behavior the "planning fallacy" and they argue that it stems from actors taking an "inside view" focusing on the constituents of the specific planned action rather than on the outcomes of similar actions that have already been completed. Kahneman and Tversky (1979b) argue that the prevalent tendency to underweigh or ignore distributional information is perhaps the major source of error in forecasting. "The analysts should therefore make every effort to frame the forecasting problem so as to facilitate utilizing all the distributional information that is available," say Kahneman and Tversky (1979b: 316). This may be considered the single most important piece of advice regarding how to increase accuracy in forecasting through improved methods. Using such distributional information from other ventures similar to that being forecasted is called taking an "outside view" and it is the cure to the planning fallacy. Reference class forecasting is a method for systematically taking an outside view on planned actions.

More specifically, reference class forecasting for a particular project requires the following three steps:

(1) Identification of a relevant reference class of past, similar projects. The class must be broad enough to be statistically meaningful but narrow enough to be truly comparable with the specific project.
(2) Establishing a probability distribution for the selected reference class. This requires access to credible, empirical data for a sufficient number of projects within the reference class to make statistically meaningful conclusions.
(3) Comparing the specific project with the reference class distribution, in order to establish the most likely outcome for the specific project.

Thus reference class forecasting does not try to forecast the specific uncertain events that will affect the particular project, but instead places the project in a statistical distribution of outcomes from the class of reference projects. In statisticians vernacular, reference class forecasting consists of regressing forecasters' best guess toward the average of the reference class and expanding their estimate of credible interval toward the corresponding interval for the class (Kahneman and Tversky 1979b: 326).

Daniel Kahneman relates the following story about curriculum planning to illustrate how reference class forecasting works (Lovallo and Kahneman 2003: 61). Some years ago, Kahneman was involved in a project to develop a curriculum for a new subject area for high schools in Israel. The project was carried out by a team of academics and teachers. In time, the team began to discuss how long the project would take to complete. Everyone on the team was asked to write on a slip of paper the number of months needed to finish and report the project. The estimates ranged from 18 to 30 months. One of the team members--a distinguished expert in curriculum development--was then posed a challenge by another team member to recall as many projects similar to theirs as possible and to think of these projects as they were in a stage comparable to their project. "How long did it take them at that point to reach completion?", the expert was asked. After a while he answered, with some discomfort, that not all the comparable teams he could think of ever did complete their task. About 40 percent of them eventually gave up. Of those remaining, the expert could not think of any that completed their task in less than seven years, nor of any that took more than ten. The expert was then asked if he had reason to believe that the present team was more skilled in curriculum development than the earlier ones had been. The expert said no, he did not see any relevant factor that distinguished this team favorably from the teams he had been thinking about. His impression was that the present team was slightly below average in terms of resources and potential. The wise decision at this point would probably have been for the team to break up, according to Kahneman. Instead, the members ignored the pessimistic information and proceeded with the project. They finally completed the project eight years later, and their efforts went largely wasted--the resulting curriculum was rarely used.

In this example, the curriculum expert made two forecasts for the same problem and arrived at very different answers. The first forecast was the inside view; the second was the outside view, or the reference class forecast. The inside view is the one that the expert and the other team members adopted. They made forecasts by focusing tightly on the project at hand, considering its objective, the resources they brought to it, and the obstacles to its completion. They constructed in their minds scenarios of their coming progress and extrapolated current trends into the future. The resulting forecasts, even the most conservative ones, were overly optimistic. The outside view is the one provoked by the question to the curriculum expert. It completely ignored the details of the project at hand, and it involved no attempt at forecasting the events that would influence the project's future course. Instead, it examined the experiences of a class of similar projects, laid out a rough distribution of outcomes for this reference class, and then positioned the current project in that distribution. The resulting forecast, as it turned out, was much more accurate.

The contrast between inside and outside views has been confirmed by systematic research (Gilovich, Griffin, and Kahneman, 2002). The research shows that when people are asked simple questions requiring them to take an outside view, their forecasts become significantly more accurate. For example, a group of students enrolling at a college were asked to rate their future academic performance relative to their peers in their major. On average, these students expected to perform better than 84% of their peers, which is logically impossible. The forecasts were biased by overconfidence. Another group of incoming students from the same major were asked about their entrance scores and their peers' scores before being asked about their expected performance. This simple diversion into relevant outside-view information, which both groups of subjects were aware of, reduced the second group's average expected performance ratings by 20%. That is still overconfident, but it is much more realistic than the forecast made by the first group (Lovallo and Kahneman 2003: 61).

However, most individuals and organizations are inclined to adopt the inside view in planning new projects. This is the conventional and intuitive approach. The traditional way to think about a complex project is to focus on the project itself and its details, to bring to bear what one knows about it, paying special attention to its unique or unusual features, trying to predict the events that will influence its future. The thought of going out and gathering simple statistics about related projects seldom enters a manager's mind. This is the case in general, according to Lovallo and Kahneman (2003: 61-62). And it is certainly the case for cost and demand forecasting in transportation infrastructure projects. Of the several hundred forecasts reviewed in Flyvbjerg, Bruzelius, and Rothengatter (2003) and Flyvbjerg, Holm, and Buhl (2002, 2005), not one was a reference class forecast.[ii]

While understandable, project managers' preference for the inside view over the outside view is unfortunate. When both forecasting methods are applied with equal skill, the outside view is much more likely to produce a realistic estimate. That is because it bypasses cognitive and political biases such as optimism bias and strategic misrepresentation and cuts directly to outcomes. In the outside view project managers and forecasters are not required to make scenarios, imagine events, or gauge their own and others' levels of ability and control, so they cannot get all these things wrong. Human bias is bypassed. Surely the outside view, being based on historical precedent, may fail to predict extreme outcomes, that is, those that lie outside all historical precedents. But for most projects, the outside view will produce more accurate results. In contrast, a focus on inside details is the road to inaccuracy.

The comparative advantage of the outside view is most pronounced for non-routine projects, understood as projects that managers and decision makers in a certain locale or organization have never attempted before--like building new plants or infrastructure or catering to new types of demand. It is in the planning of such new efforts that the biases toward optimism and strategic misrepresentation are likely to be largest. To be sure, choosing the right reference class of comparative past projects becomes more difficult when managers are forecasting initiatives for which precedents are not easily found, for instance the introduction of new and unfamiliar technologies. However, most projects are both non-routine locally and use well-known technologies. Such projects are, therefore, particularly likely to benefit from the outside view and reference class forecasting.

**First Instance of Reference Class Forecasting in Practice**

The first instance of reference class forecasting in practice may be found in Flyvbjerg and Cowi (2004): *Procedures for Dealing with Optimism Bias in Transport Planning.*[iiii] Based on this study, in the Summer of 2004 the UK Department for Transport and HM Treasury decided to employ the method as part of project appraisal for large transportation projects.

The immediate background to this decision was the revision to "The Green Book" by HM Treasury in 2003, that identified for large public procurement a demonstrated, systematic tendency for project appraisers to be overly optimistic:

> "There is a demonstrated, systematic, tendency for project appraisers to be overly optimistic. To redress this tendency appraisers should make explicit, empirically based adjustments to the estimates of a project's costs, benefits, and duration ... [I]t is recommended that these adjustments be based on data from past projects or similar projects elsewhere" (HM Treasury 2003b: 1).

Such optimism was seen as an impediment to prudent fiscal planning, for the government as a whole and for individual departments within government. To redress this tendency HM Treasury recommended that appraisers involved in large public procurement should make explicit, empirically based adjustments to the estimates of a project's costs, benefits, and duration. HM Treasury recommended that these adjustments be based on data from past projects or similar projects elsewhere, and adjusted for the unique characteristics of the project at hand. In the absence of a more specific evidence base, HM Treasury encouraged government departments to collect valid and reliable data to inform future estimates of optimism, and in the meantime use the best available data. The Treasury let it be understood that in future the allocation of funds for large public procurement would be dependent on valid adjustments of optimism in order to secure valid estimates of costs, benefits, and duration of large public procurement (HM Treasury 2003a, b).

In response to the Treasury's Green Book and its recommendations, the UK Department for Transport decided to collect the type of data, which the Treasury recommended, and on that basis to develop a methodology for dealing with optimism bias in the planning and management of transportation projects. The Department for Transport appointed Bent Flyvbjerg in association with Cowi to undertake this assignment as regards costing of large transportation procurement. The main aims of the assignment were two; first, to provide empirically based optimism bias uplifts for selected reference classes of transportation infrastructure projects, and, second, to provide guidance on using the established uplifts to produce more realistic forecasts of capital expenditures in individual projects (Flyvbjerg and Cowi 2004). Uplifts would be established for capital expenditures based on the full business case (time of decision to build).

The types of transportation schemes under the direct and indirect responsibility of the UK Department for Transport were divided into a number of distinct categories where statistical tests, benchmarkings, and other analyses showed that the risk of cost overruns within each category may be treated as statistically similar. For each category a reference class of projects was then established as the basis for reference class forecasting, as required by step 1 in the 3-step procedure for reference class forecasting described above. The specific categories and the types of project allocated to each category are shown in Table 3.

For each category of projects, a reference class of completed, comparable transportation infrastructure projects was used to establish probability distributions for cost overruns for new projects similar in scope and risks to the projects in the reference class, as required by step 2 in reference class forecasting. For roads, for example, a class of 172 completed and comparable projects was used to establish the probability distribution of cost overruns shown in Figure 2. The share of projects with a given maximum cost overrun is shown in the figure. For instance, 40% of projects have a maximum cost overrun of 10%; 80% of projects a maximum overrun of 32%, etc. For rail, the probability distribution is shown in Figure 3, and for bridges and tunnels in Figure 4. The figures show that the risk of cost overrun is substantial for all three project types, but highest for rail, followed by bridges and tunnels, and with the lowest risk for roads.

Based on the probability distributions described above the required uplifts needed to carry out step 3 in a reference class forecast may be calculated as shown in Figures 5-7. The uplifts refer to cost overrun calculated in constant prices. The lower the acceptable risk for cost overrun, the higher the uplift. For instance, with a willingness to accept a 50% risk for cost overrun in a road project, the required uplift for this project would be 15%. If the Department for Transport were willing to accept only a 10% risk for cost overrun, then the required uplift would be 45%. In comparison, for rail with a willingness to accept a 50% risk for cost overrun, the required uplift would be 40%. If the Department for Transport were willing to accept only a 10% risk for cost overrun, then the required uplift would be 68% for rail. All three figures share the same basic S-shape, but at different levels, demonstrating that the required uplifts are significantly different for different project categories for a given level of risk of cost overrun. The figures also show that the cost for additional reductions in the risk of cost overrun is different for the three types of projects, with risk reduction becoming increasingly expensive (rising marginal costs) for roads and fixed links below 20% risk, whereas for rail the cost of increased risk reduction rises more slowly, albeit from a high level.

Table 4 presents an overview of applicable optimism bias uplifts for the 50% and 80% percentiles for all the project categories listed in Table 3. The 50% percentile is pertinent to the investor with a large project portfolio, where cost overruns on one project may be offset by cost savings on another. The 80% percentile--corresponding to a risk of cost overrun of 20%--is the level of risk that the UK Department for Transport is typically willing to accept for large investments in local transportation infrastructure.

The established uplifts for optimism bias should be applied to estimated budgets at the time of decision to build a project. In the UK, the approval stage for a large transportation project is equivalent to the time of presenting the business case for the project to the Department for Transport with a view to obtaining the go or no-go for that project.

If, for instance, a group of project managers were preparing the business case for a new motorway, and if they or their client had decided that the risk of cost overrun must be less than 20%, then they would use an uplift of 32% on their estimated capital expenditure budget. Thus, if the initially estimated budget were £100 million, then the final budget--taking into account optimism bias at the 80%-level--would be £132 million (£1 = $1.8). If the project managers or their client decided instead that a 50% risk of cost overrun was acceptable, then the uplift would be 15% and the final budget £115 million.

Similarly, if a group of project managers were preparing the business case for a metro rail project, and if they or their client had decided that with 80% certainty they wanted to stay within budget, then they would use an uplift on capital costs of 57%. An initial capital expenditure budget of £300 million would then become a final budget of £504 million. If the project managers or their client required only 50% certainty they would stay within budget, then the final budget would be £420 million.

It follows that the 50% percentile should be used only in instances where investors are willing to take a high degree of risk that cost overrun will occur and/or in situations where investors are funding a large number of projects and where cost savings (underruns) on one project may be used to cover the costs of overruns on other projects. The upper percentiles (80-90%) should be used when investors want a high degree of certainty that cost overrun will not occur, for instance in stand-alone projects with no access to additional funds beyond the approved budget. Other percentiles may be employed to reflect other degrees of willingness to accept risk and the associated uplifts as shown in Figures 5-7.

Only if project managers have evidence to substantiate that they would be significantly better at estimating costs for the project at hand than their colleagues were for the projects in the reference class would the managers be justified in using lower uplifts than those described above. Conversely, if there is evidence that the project managers are worse at estimating costs than their colleagues, then higher uplifts should be used.

The methodology described above for systematic, practical reference class forecasting for transportation projects was developed 2003-2004 with publication by the Department of Transport in August 2004. From this date local authorities applying for funding for transportation projects with the Department for Transport or with HM Treasury were required to take into account optimism bias by using uplifts as described above and as laid out in more detail in guidelines from the two ministries.

# Forecasting Costs for the Edinburgh Tram

In October 2004, the first instance of practical use of the uplifts was recorded, in the planning of the Edinburgh Tram Line 2. Ove Arup and Partners Scotland (2004) had been appointed by the Scottish Parliament's Edinburgh Tram Bill Committee to provide a review of the Edinburgh Tram Line 2 business case developed on behalf of Transport Initiatives Edinburgh. Transport Initiatives Edinburgh is project promoter and is a private limited company owned by the City of Edinburgh Council established to deliver major transport projects for the Council. The Scottish Executive is a main funder of the Edinburgh Tram, having made an Executive Grant of £375 million ($670 million) towards lines 1 and 2 of which Transport Initiatives Edinburgh proposed spending £165 million towards Line 2.

As part of their review, Ove Arup assessed whether the business case for Tram Line 2 had adequately taken into account optimism bias as regards capital costs. The business case had estimated a base cost of £255 million and an additional allowance for contingency and optimism bias of £64 million--or 25%--resulting in total capital costs of app. £320 million. Ove Arup concluded about this overall estimate of capital costs that it seemed to have been rigorously prepared using a database of costs, comparison to other UK light rail schemes, and reconciliations with earlier project estimates. Ove Arup found, however, that the following potential additional costs needed to be considered in determining the overall capital costs: £26 million for future expenditure on replacement and renewals and £20 million as a notional allowance for a capital sum to cover risks of future revenue shortfalls, amounting to an increase in total capital costs of 14.4% (Ove Arup and Partners Scotland 2004: 15-16)

Using the UK Department for Transport uplifts for optimism bias presented above on the base costs, Ove Arup then calculated the $80^{th}$ percentile value for total capital costs--the value at which the likelihood of staying within budget is 80%--to be £400 million (i.e., £255 million x 1.57). The $50^{th}$ percentile for total capital costs--the value at which the likelihood of staying within budget is 50%--was £357 million (i.e., £255 x 1.4). Ove Arup remarked that these estimates of total capital costs were likely to be conservative, that is, low, because the UK Department for Transport recommends that its optimism bias uplifts be applied to the budget at the time of decision to build, which typically equates to business case submission, and Tram Line 2 had not yet even reached the outline business case stage, indicating that risks would be substantially higher at this early stage as would corresponding uplifts. On that basis Arup concluded that "it is considered that current optimism bias uplifts [for Tram Line 2] may have been underestimated" (Ove Arup and Partners Scotland 2004: 27).

Finally, Ove Arup mentioned that the Department for Transport guidance does allow for optimism bias to be adjusted downward if strong evidence of improved risk mitigation can be demonstrated. According to Ove Arup, this may be the case if advanced risk analysis has been applied, but this was not the case for Tram Line 2. Ove Arup therefore concluded that "the justification for reduced Department for Transport optimism bias uplifts would appear to be weak" (Ove Arup and Partners Scotland 2004: 27-28). Thus the overall conclusion of Ove Arup was that the promoter's capital cost estimate of app. £320 million was optimistic. Most likely Tram Line 2 would cost significantly more.

By framing the forecasting problem to allow the use of the empirical distributional information made available by the UK Department for Transport, Ove Arup was able to take an outside view on the Edinburgh Tram Line 2 capital cost forecast and thus debias what appeared to be a biased forecast. As a result Ove Arup's client, The Scottish Parliament, was provided with a more reliable estimate of what the true costs of Line 2 was likely to be.

## Potentials and Barriers for Reference Class Forecasting

As mentioned above, two types of explanation best account for forecasting inaccuracy, optimism bias and strategic misrepresentation. Reference class forecasting was originally developed to mitigate optimism bias, but reference class forecasting may help mitigate any type of human bias, including strategic bias, because the method bypasses such bias by cutting directly to empirical outcomes and building forecasts on these. Even so, the potentials for and barriers to reference class forecasting will be different in situations where (1) optimism bias is the main cause of inaccuracy as compared to situations where (2) strategic misrepresentation is the reason for inaccuracy. We therefore need to distinguish between these two types of situation when endeavoring to apply reference class forecasting in practice.

In the first type of situation--where optimism bias is the main cause of inaccuracy--we may assume that managers and forecasters are making honest mistakes and have an interest in improving accuracy. Consider, for example, the students mentioned above, who were asked to estimate their future academic performance relative to their peers. We may reasonably believe that the students did not deliberately misrepresent their estimates, because they had no interest in doing so and were not exposed to pressures that would push them in that direction. The students made honest mistakes, which produced honest, if biased, numbers regarding performance. And, indeed, when students were asked to take into account outside-view information, we saw that the accuracy of their estimates improved substantially. In this type of situation--when forecasters are honestly trying to gauge the future--the potential for using the outside view and reference class forecasting will be good. Forecasters will be welcoming the method and barriers will be low, because no one has reason to be against a methodology that will improve their forecasts.

In the second type of situation--where strategic misrepresentation is the main cause of inaccuracy--differences between estimated and actual costs and benefits are best explained by political and organizational pressures. Here managers and forecasters would still need reference class forecasting if accuracy were to be improved, but managers and forecasters may not be interested in this because inaccuracy is deliberate. Biased forecasts serve strategic purposes that dominate the commitment to accuracy and truth. Consider, for example, city managers with responsibility for estimating costs and benefits of urban rail projects. Here, the assumption of innocence regarding outcomes typically cannot be upheld. Cities compete fiercely for approval and for scarce national funds for such projects, and pressures are strong to present projects as favorably as possible, that is, with low costs and high benefits, in order to beat the competition. There is no incentive for the individual city to debias its forecasts, but quite the opposite. Unless all other cities also debias, the individual city would lose out in the competition for funds. Project managers are on record confirming that this is a common situation (Flyvbjerg and Cowi 2004: 36-58; Flyvbjerg and Lovallo in progress). The result is the same as in the case of optimism: actors promote ventures that are unlikely to perform as promised. But the causes are different as are possible cures.

In this type of situation the potential for reference class forecasting is low--the demand for accuracy is simply not there--and barriers are high. In order to lower barriers, and thus create room for reference class forecasting, measures of accountability must be implemented that would reward accurate forecasts and punish inaccurate ones.

Forecasters and promoters should be made to carry the full risks of their forecasts. Their work should be reviewed by independent bodies such as national auditors or independent analysts, and such bodies would need reference class forecasting to do their work. Projects with inflated benefit-cost ratios should be stopped or placed on hold. Professional and even criminal penalties should be considered for people who consistently produce misleading forecasts. The higher the stakes, and the higher the level of political and organizational pressures, the more pronounced will be the need for such measures of accountability. Flyvbjerg, Bruzelius, and Rothengatter (2003) and Flyvbjerg, Holm, and Buhl (2005) further detail the design of such measures and how they may be implemented in practical project management.

The existence of strategic misrepresentation does not exclude the simultaneous existence of optimism bias, and vice versa. In fact, it is realistic to expect such co-existence in forecasting in large and complex projects and organizations. This again underscores the point that improved forecasting methods--here reference class forecasting--and measures of accountability must go hand in hand if the attempt to arrive at more accurate forecasts is to be effective.

Finally, it could be argued that in some cases the use of reference class forecasting may result in such large reserves set aside for a project that this would in itself lead to risks of inefficiencies and overspending. Reserves will be spent simply because they are there, as the saying goes in the construction business. For instance, it is important to recognize that for the abovementioned examples the introduction of reference class forecasting and optimism-bias uplifts would establish total budget reservations (including uplifts) which for some projects would be more than adequate. This may in itself create an incentive which works against firm cost control if the total budget reservation is perceived as being available to the project and its contractors. This makes it important to combine the introduction of reference class forecasting and optimism bias uplifts with tight contracts, maintained incentives for promoters to undertake good quantified risk assessment and exercise prudent cost control during project implementation. How this may be done is described in Flyvbjerg and Cowi (2004).

18/32

| Type of project | Average inaccuracy (%) | Standard deviation | Level of significance, p |
|---|---|---|---|
| Rail | 44.7 | 38.4 | <0.001 |
| Bridges and tunnels | 33.8 | 62.4 | 0.004 |
| Road | 20.4 | 29.9 | <0.001 |

**Table 1: Inaccuracy in cost forecasts for rail, bridges, tunnels, and roads, respectively (construction costs, constant prices). For all project types inaccuracy is different from zero with extremely high significance.**
**Source: Flyvbjerg database on large-scale infrastructure projects.**

|  | **Rail** | **Road** |
|---|---|---|
| Average inaccuracy (%) | -51.4 (sd=28.1) | 9.5 (sd=44.3) |
| Percentage of projects with inaccuracies larger than ±20% | 84 | 50 |
| Percentage of projects with inaccuracies larger than ±40% | 72 | 25 |
| Percentage of projects with inaccuracies larger than ±60% | 40 | 13 |

**Table 2: Inaccuracy in forecasts of rail passenger and road vehicle traffic.**
**Source: Flyvbjerg database on large-scale infrastructure projects.**

| **Category** | **Types of projects** |
|---|---|
| Roads | Motorway |
| | Trunk roads |
| | Local roads |
| | Bicycle facilities |
| | Pedestrian facilities |
| | Park and ride |
| | Bus lane schemes |
| | Guided buses on wheels |
| Rail | Metro |
| | Light rail |
| | Guided buses on tracks |
| | Conventional rail |
| | High speed rail |
| Fixed links | Bridges |
| | Tunnels |
| Building projects | Stations |
| | Terminal buildings |
| IT projects | IT system development |
| Standard civil engineering | Included for reference purposes only |
| Non-standard civil engineering | Included for reference purposes only |

**Table 3: Categories and types of projects used as basis for reference class forecasting**

| Category | Types of projects | Applicable optimism bias uplifts | |
|---|---|---|---|
| | | **50% percentile** | **80% percentile** |
| Roads | Motorway<br>Trunk roads<br>Local roads<br>Bicycle facilities<br>Pedestrian facilities<br>Park and ride<br>Bus lane schemes<br>Guided buses on wheels | 15% | 32% |
| Rail | Metro<br>Light rail<br>Guided buses on tracks<br>Conventional rail<br>High speed rail | 40% | 57% |
| Fixed links | Bridges<br>Tunnels | 23% | 55% |
| Building projects | Stations<br>Terminal buildings | 4-51%* | |
| IT projects | IT system development | 10-200%* | |
| Standard civil engineering | Included for reference purposes only | 3-44%* | |
| Non-standard civil engineering | Included for reference purposes only | 6-66%* | |

*) Based on Mott MacDonald (2002: 32); no probability distribution available.

**Table 4: Applicable capital expenditure optimism bias uplifts for 50% and 80% percentiles, constant prices.**

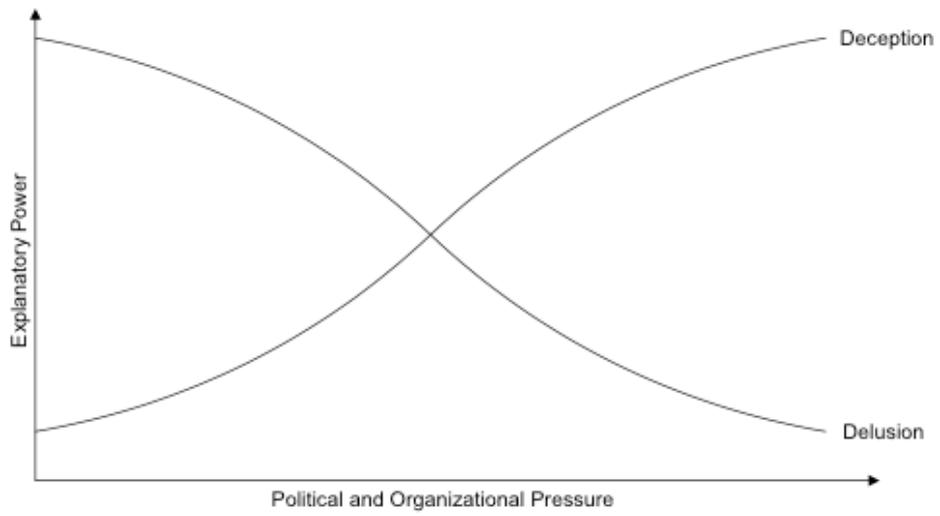

**Figure 1: Explanatory power of optimism bias and strategic misrepresentation, respectively, in accounting for forecasting inaccuracy as function of political and organizational pressure.**

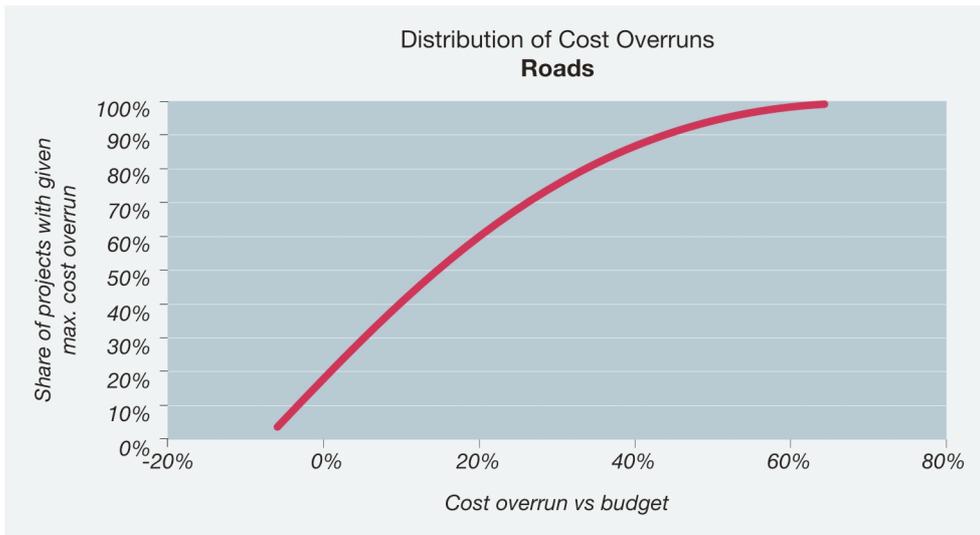

**Figure 2: Probability distribution of cost overrun for roads, constant prices (N=172).**
**Source: Flyvbjerg database on large-scale infrastructure projects.**

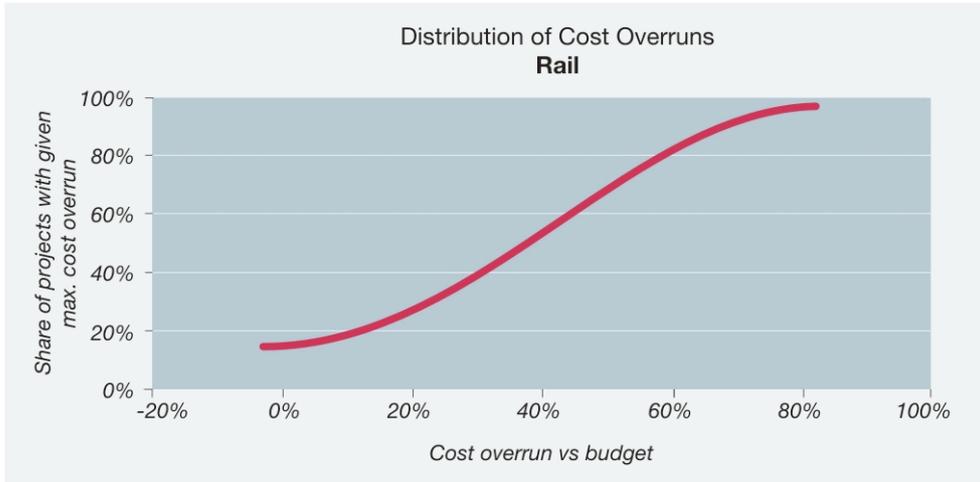

**Figure 3: Probability distribution of cost overrun for rail, constant prices (N=46).**
**Source: Flyvbjerg database on large-scale infrastructure projects.**

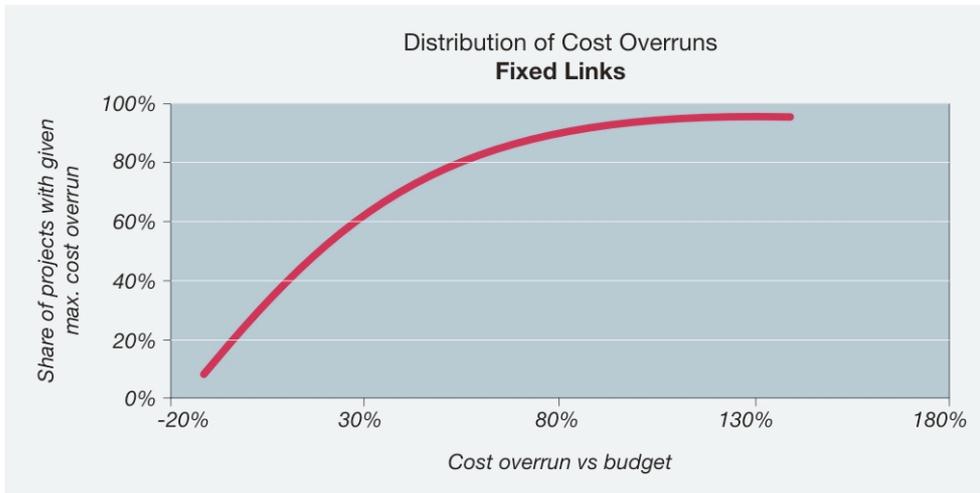

**Figure 4: Probability distribution of cost overrun for fixed links, constant prices (N=34).**
**Source: Flyvbjerg database on large-scale infrastructure projects.**

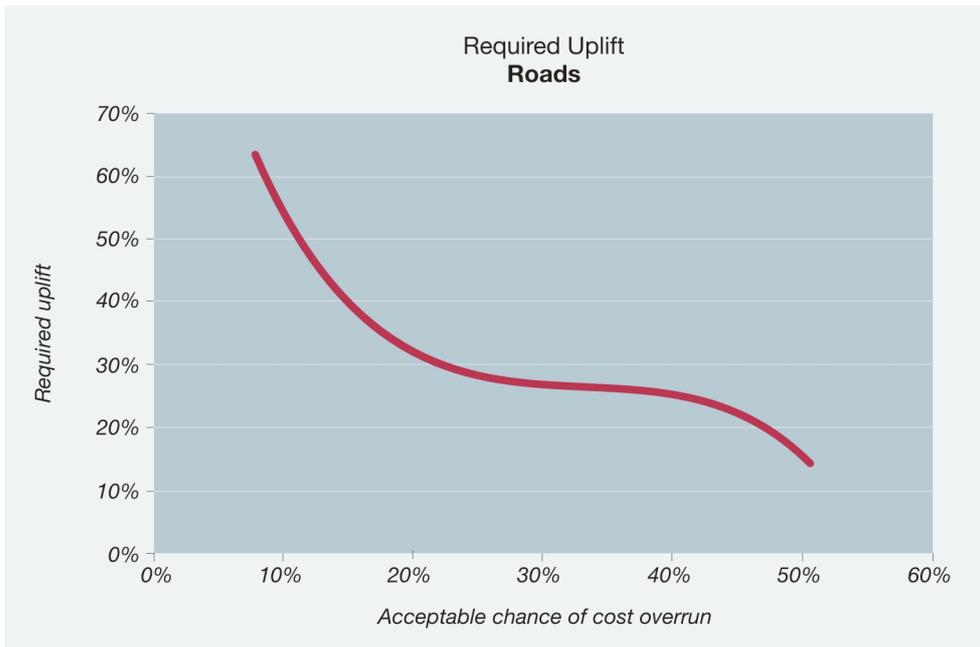

**Figure 5: Required uplift for roads as function of the maximum acceptable level of risk for cost overrun, constant prices (N=172).**
**Source: Flyvbjerg database on large-scale infrastructure projects.**

30/32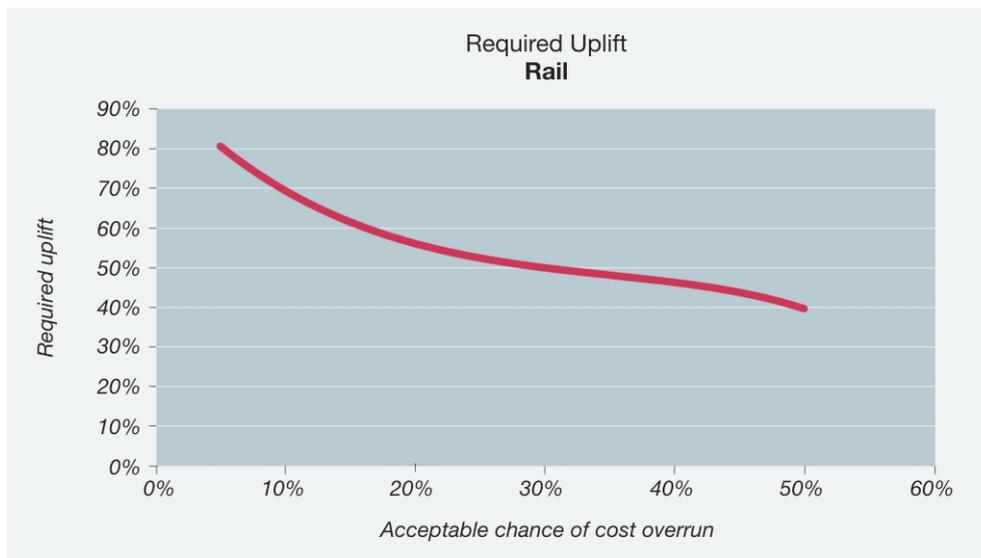

**Figure 6: Required uplift for rail as function of the maximum acceptable level of risk for cost overrun, constant prices (N=46).**
**Source: Flyvbjerg database on large-scale infrastructure projects.**

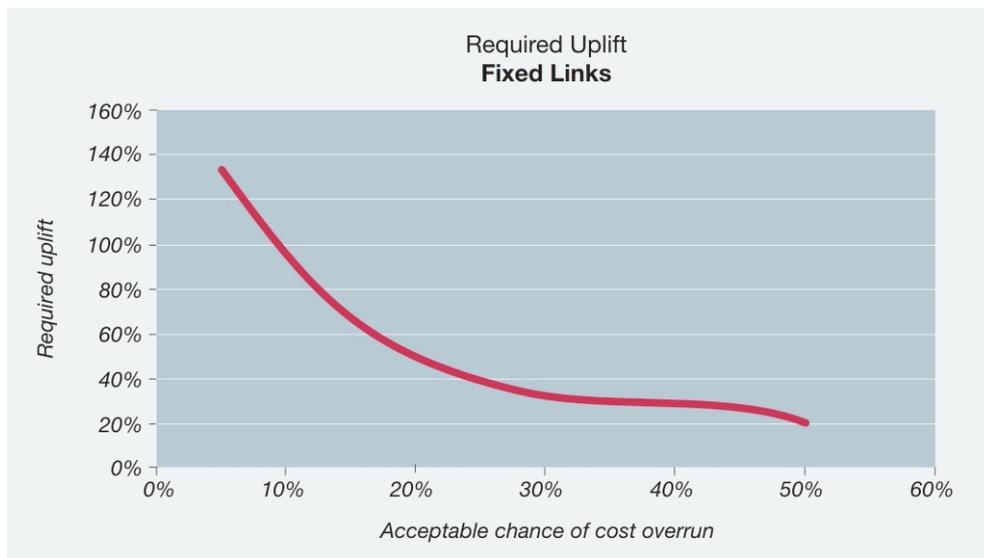

**Figure 7: Required uplift for fixed links as function of the maximum acceptable level of risk for cost overrun, constant prices (N=34).**
**Source: Flyvbjerg database on large-scale infrastructure projects.**

# Notes

[i] Inaccuracy is measured in percent as (actual outcome/forecast outcome -1)x100. The base year of a forecast for a project is the time of decision to build that project. An inaccuracy of 0 indicates perfect accuracy. Cost is measured as construction costs. Demand is measured as number of vehicles for roads and number of passengers for rail.

[ii] The closest thing to an outside view in large infrastructure forecasting is Gordon and Wilson's (1984) use of regression analysis on an international cross section of light-rail projects to forecast patronage in a number of light-rail schemes in North America.

[iii] The fact that this is, indeed, the first instance of practical reference class forecasting has been confirmed with Daniel Kahneman and Dan Lovallo, who also knows of no other instances of practical reference class forecasting. Personal communications with Daniel Kahneman and Dan Lovallo, author's archives.